\documentclass[12pt,preprint]{aastex}

\shorttitle{Water masers in planetary nebulae}
\shortauthors{de Gregorio-Monsalvo et al.}

\begin{document}

\title{A survey for water maser emission towards planetary nebulae. 
New detection in IRAS 17347-3139}

\author{Itziar de Gregorio-Monsalvo\altaffilmark{1,2}, 
  Yolanda G\'omez\altaffilmark{3}, Guillem
  Anglada\altaffilmark{4}, Riccardo Cesaroni\altaffilmark{5}, Luis
  F. Miranda\altaffilmark{4}, Jos\'e
  F. G\'omez\altaffilmark{1}, Jos\'e M. Torrelles\altaffilmark{6}}

\altaffiltext{1}{Laboratorio de Astrof\'{\i}sica Espacial y F\'{\i}sica
  Fundamental, INTA, Apartado 50727, E-28080
    Madrid, Spain; e-mail: itziar@laeff.esa.es; jfg@laeff.esa.es}
\altaffiltext{2}{National Radio Astronomy Observatory, PO
Box O, Socorro, NM 87801}
\altaffiltext{3}{Centro de Radioastronom\'{\i}a y Astrof\'{\i}sica,
  UNAM, Apdo. Postal 3-72 (Xangari), 58089 Morelia, Michoac\'an, M\'exico} 
\altaffiltext{4}{Instituto de Astrof\'{\i}sica de Andaluc\'{\i}a
  (CSIC), Apartado 3004, E-18080 Granada, Spain}
\altaffiltext{5}{INAF, Osservatorio
         Astrofisico di Arcetri, Largo E. Fermi 5, 50125 Firenze, Italy}
\altaffiltext{6}{Instituto de Ciencias del Espacio (CSIC)-IEEC, Gran
Capit\`a 2, E-08034 Barcelona, Spain}

\begin{abstract}
We report on a water maser survey towards a sample of 27 planetary
nebulae (PNe) using the 
Robledo de Chavela and Medicina single-dish antennas, as
well as the Very Large Array (VLA). Two detections have been obtained: 
the already known water maser emission
in  K 3-35, and a new cluster of masers in IRAS 17347-3139. This low rate of
detections is compatible with the short life-time of water molecules
in PNe ($\sim 100$ yr). The water maser cluster at IRAS 17347-3139 are distributed
on a ellipse of size $\simeq 0\farcs 2\times 0\farcs 1$, spatially
associated with compact 1.3 cm continuum emission (simultaneously
observed with the VLA). From archive VLA continuum data at 4.9, 8.4, and
14.9 GHz, 
a spectral index $\alpha = 0.76 \pm 0.03$ ($S_\nu \propto
\nu^{\alpha}$) is derived for this radio source,
which is consistent
with either a partially optically thick ionized region or with
an ionized wind. 
However, the latter scenario can be ruled out on mass-loss
considerations, thus indicating that this source is probably a young PN.
The spatial distribution and the radial
velocities of the water masers are suggestive of a rotating and
expanding maser ring,
tracing the innermost regions of a torus formed at the end of the AGB phase.
Given that the 1.3 cm continuum emission peak is located near one of the
tips of the major axis of the ellipse of masers, we speculate
on a possible binary nature of IRAS 17347-3139,
where the radio continuum emission could belong to one of the components and the
water masers would be associated with a companion.
\end{abstract}


\section{Introduction}

Nowadays we know that planetary nebulae (PNe) are the result of the
evolution of  carbon- and 
oxygen-rich red giants, after going through the AGB and 
a brief ($\simeq$ 1000 yr) proto-PN phase
\citep{Kwo93}. 
In a simplified picture, a PN forms when the remnant 
(central) star reaches an effective temperature (T$_{\rm eff}$) of $\sim$ 
30\,000\,K and the envelope ejected during the AGB phase is ionized. 
However, the transformation of an AGB star into a PN is much more complex. The 
spherical symmetry typical of AGB envelopes is lost in the transition, 
and most PNe exhibit elliptical and bipolar shells \citep{Bal87}. In 
addition, highly collimated outflows develop in the proto-PN and/or PN
phase, which may play a crucial role in the dynamical and shaping processes 
of PNe \citep{SyT98}. 

Envelopes of evolved stars with oxygen-rich chemistry are characterized by 
strong OH, H$_2$O and SiO maser emission \citep{RyM81,Eli92,Hab96}. 
Because of its high intensity
and the possibility to be observed at high spatial and spectral resolution, 
maser emission is ideal to study dynamical and physical conditions of 
circumstellar envelopes \citep{Hab96,Eng02}. Masers 
are stratified in these envelopes with SiO masers located close to the 
stellar surface, 
water masers at about 10--100 AU, and OH masers further away, up to
$\sim 10^4$ AU \citep{lane87}. 
In the 
transformation of an AGB star into a PN, masers disappear sequentially 
\citep{Lew89}. When pulsation and mass loss cease at the tip of the AGB, 
SiO masers 
disappear \citep{Gom90,Nym98,Eng02}, while water and OH masers can still be 
found in the proto-PN phase \citep {Lik88,MyB99,Sah99a,Sah99b}.  
As the star enters its PN phase, water molecules are rapidly destroyed by 
the ionizing radiation \citep{Lew89,Gom90}. Thus, 
water masers are not expected in PNe, and only OH masers seem to 
persist at the 
very early stages of their evolution.  

Surprisingly, water maser emission has recently been found in the PN
K 3-35 (Miranda et al. 2001, hereafter MGAT01). K 3-35 is a young 
bipolar PN that contains a bright compact core, a dense equatorial 
torus and a bipolar precessing jet, with all these structures surrounded 
by a faint elliptical envelope \citep{AyK89,Aaq93,Mir98,Mir00,Mir01}. 
Water maser emission was detected in two different regions of 
K 3-35: in the torus, at a distance of about $\simeq$ 85 AU from the
center
(assuming a 
distance of 5 kpc to K 3-35, Zhang 1995), and at the tips 
of the bipolar radio jet, 
at $\simeq$ 5,000 AU from the central star. The time scale over which water 
molecules are expected to survive destruction by the ionization front is very short, 
$\sim 100$ years \citep{Gom90}. Thus, it was inferred that 
K 3-35 had been caught at the very moment of its transformation into a 
PN (MGAT01). In addition, water maser emission was not expected so 
far away from the central star, because of the lack of physical conditions to 
provide the energy to pump the water maser \citep{Mar97}.  The bipolar jet 
in K 3-35 was proposed as the excitation 
agent of the distant water masers, because of the
former's spatial and kinematic association with the latter. 
For this source, some unknown and local shielding 
mechanism against 
ionizing radiation was invoked to explain why water molecules still
survive. Some recent models of gas-phase chemistry, however, predict the 
production of water and other molecules in the envelopes of PN, long
after the original molecules were photodissociated \citep{Ali01, Has00}.

To better understand the phenomenon of water emission in 
PNe, and to determine whether K 3-35 is representative of this kind of 
sources, or an extremely rare case, we have undertaken a survey of water maser 
emission towards PNe. In this paper we present the results of our
survey. A technical description of the observations is presented in
\S2. Our results are presented in \S3 and their discussion in
\S4. Finally, we summarize our conclusions in \S5.

\section{Observations}

We observed the  6$_{16} \rightarrow$5$_{23}$ transition of the water molecule 
(rest frequency 22235.080 MHz) towards a sample of planetary nebulae
(see Table \ref{tbsurvey}), using
the NASA 70-m antenna (DSS-63) at Robledo de Chavela (Spain), the 32-m antenna\footnote{The Medicina
VLBI radio telescope is operated by the Radioastronomy Institute of the
Italian National Research Council (CNR)}
at Medicina (Italy), and the Very Large Array (VLA) of the National Radio 
Astronomy Observatory\footnote{The National Radio Astronomy Observatory
is a facility of the National Science Foundation operated under
cooperative agreement by Associated Universities, Inc.} (USA). Several sources were observed with more
than one telescope. Further simultaneous water maser and 1.3 cm
continuum observations were made with the VLA towards IRAS 17347-3139,
the source where maser emission was first detected  
at Robledo de Chavela. We also analyzed VLA 
archive continuum data of this source. Technical details of these
observations are described below.

The observed sources were chosen with the criterion of showing some
sign of being associated with molecular gas and/or dust, e.g.,
CO or OH lines,
or (far or near) infrared emission.

\subsection{Robledo de Chavela 70-m}

The 1.3 cm receiver of this antenna was a cooled HEMT. 
Water maser observations were carried out using a 
256-channel autocorrelation spectrometer, covering a bandwidth of 10
MHz, which provides a velocity
resolution of 0.5 km s$^{-1}$. At this frequency, the half-power beamwidth
of the telescope is $\sim 41''$. Spectra were taken 
in frequency switching mode, with a switch of 5 MHz, thus providing an
effective
velocity coverage of 202.4 km s$^{-1}$ (15 MHz) centered at the $V_{\rm 
LSR}$ of each source. Only left circular polarization was processed.
Calibration
was performed using a noise diode. 
These observations were carried out between 2002 April and
August, with system temperatures between 45 and 135 K, and a total integration
time of 20 min per source. The rms pointing accuracy was
better than 15$''$. A total of 15 sources were observed with this
antenna (see Table \ref{tbsurvey}). The data reduction was performed using
the CLASS package, which is part of the GAG software package 
developed at IRAM and
Observatoire de Grenoble.

\subsection{Medicina 32-m} 

Observations with the Medicina 32-m antenna were carried out in
several periods from March to October 2002, making use of a digital
autocorrelator spectrometer with a bandwidth of 8~MHz
(108~km~s$^{-1}$ at the frequency of the H$_2$O 6$_{16}$--5$_{23}$ line)
and a spectral resolution of 9.77~kHz.
A GaAs FET amplifier was used as a front-end, which allows for zenith system
temperatures of $\sim$120~K under good weather conditions. During the
observations the system temperature varied from 130 to 290~K depending on
weather conditions and elevation. Observations were made in total power mode,
with integrations of 5~min on- and 5~min off-source. 
Then the observation was repeated as
many times as needed to reach typical 1$\sigma$ rms noise levels of
$\sim$0.5~Jy. For each source, all spectra taken on the same day were
averaged and smoothed to a spectral resolution of 0.263~km~s$^{-1}$. The
antenna gain as a function of elevation was determined by total power
integrations on DR21 (adopted flux density 18.8~Jy) repeated every
1--2~hours. Pointing was checked on strong water maser sources and the
resulting accuracy is 20$''$. Nine sources were observed with this
antenna. The data reduction was performed with
CLASS.

\subsection{Very Large Array}

The VLA observations were made in two different periods. As part of our 
water maser survey, on 2002 May 5 we
observed 14 planetary nebulae 
in the A configuration. The integration
time for each source 
was $\sim$ 15 minutes. We sampled 63 channels with 97.7 kHz resolution 
($\sim$1.3 km s$^{-1}$ at the observed frequency), in both right and left circular polarizations. 
The source \objectname{3C 48} was used as the primary 
calibrator for flux density, for which we adopted a flux density of 
1.13 Jy using
the latest VLA values (1999.2). Amplitude calibration was performed 
using the model of 3C 48 downloaded from the NRAO web site. 
Phase calibrators used for each target source are listed in Table \ref{tbcals}. 
The calibration and data reduction were carried out 
using the Astronomical Image Processing System (AIPS) of NRAO. Maps
were produced setting the ``robust'' weight parameter to 0, to optimize the
tradeoff between angular resolution and sensitivity, and deconvolved
using the CLEAN algorithm.
The resulting synthesized beams are also shown in Table \ref{tbcals}.
In K 3-35, self-calibration was performed, given its strong
water maser emission. To mitigate the Gibbs ringing, spectral Hanning
smoothing was applied to all sources, 
 giving a final velocity resolution of 2.6 km s$^{-1}$.

On 2002 July 9 we made simultaneous observations of 1.3 cm continuum
and water maser towards IRAS 17347-3139 with the B configuration,
using the 4IF spectral line mode of the
VLA. For the continuum measurements, we used two of the four IFs, with
a bandwidth of 25 MHz, and seven channels 3.125 MHz wide each,
 centered at 22285.080 MHz. For the maser emission
observations (made with the other two IFs) 
we used a bandwidth of 3.125 MHz, sampled with 63 channels of
48.8 kHz each (0.66 km s$^{-1}$). Spectral Hanning smoothing was
applied, with a final velocity resolution of 1.3 km s$^{-1}$. The source
\objectname{3C 286} was used as the primary 
calibrator for flux density (adopted flux density = 2.53 Jy), while the
source J1745-290 was used as phase calibrator (bootstrapped flux density
= 0.986$\pm$0.008 Jy). The on-source integration time was $\sim$ 1 h. 
Initial calibration was done separately for each pair of IFs to remove
electronic phase differences between IFs.
Self-calibration was performed using the strongest water maser
component identified in the narrow bandwidth.
The phase and amplitude corrections were then applied to both the narrow
and broad bandwidth, removing
both atmospheric and instrumental errors.
Cleaned maps for both line and
continuum were produced by setting
the ``robust'' weight parameter to 0, resulting in a
synthesized beam of $0\farcs 81 \times 0\farcs 26$ (P.A = 26$\degr$).

Finally, we have analyzed VLA archive continuum data 
at 4.9, 8.4, and 14.9 GHz in IRAS 17347-3139 (from VLA project AP192). 
These data were observed in DnC configuration on 1991 February 24. 
Two independent IFs 
of 50 MHz bandwidth were used, providing an effective bandwidth of 100 MHz. 
Both right and left
circular polarizations were observed. 
The on-source integration time at each of the observed frequecies was
$\sim$ 10 minutes. The source 
3C 286 was used as flux calibrator, with adopted flux 
densities of
3.45 Jy (14.9 GHz), 5.21 Jy (8.4 GHz), and 7.48 Jy (4.9 GHz). The
phase calibrator was B1730-130, with bootstrapped  flux
densities of 7.02$\pm$0.20 Jy (14.9 GHz), 6.66$\pm$0.06 Jy (8.4 GHz) and
6.63$\pm$0.02 Jy (4.9 GHz). Data were self calibrated, and
cleaned images were obtained setting the ``robust'' weight parameter
to 0, resulting in 
synthesized beams of $4\farcs 24 \times 3\farcs 13$ (P.A. =
44$\degr$), $7\farcs 90 \times 5\farcs 56$ (P.A.=
40$\degr$) and $12\farcs 62 \times 10\farcs 55$ (P.A.= 61$\degr$) 
at 14.9, 8.4, and 4.9 GHz respectively.

\section{Results}

\subsection{Survey of water masers}

\begin{figure}
\rotatebox{-90}{
\plottwo{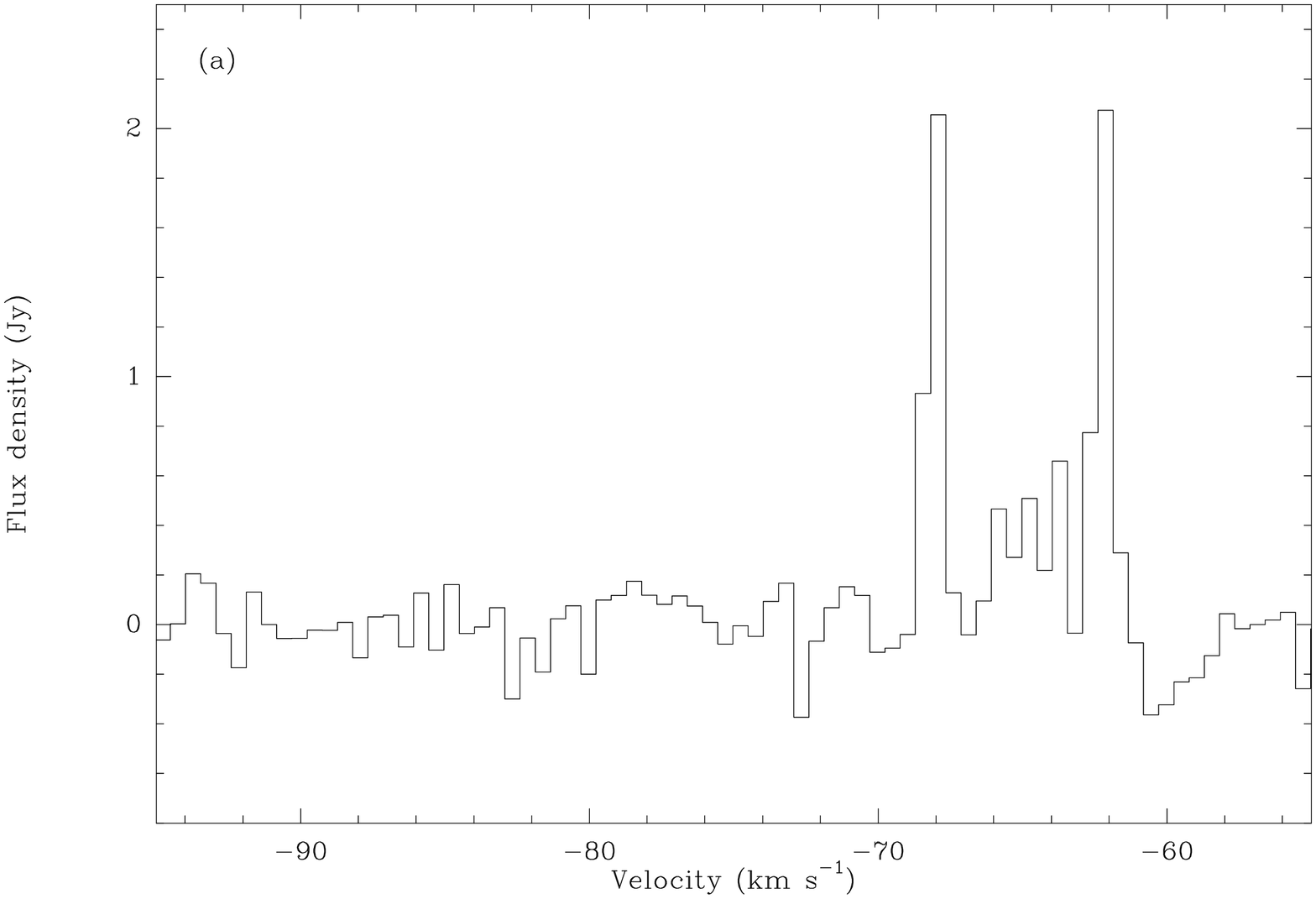}{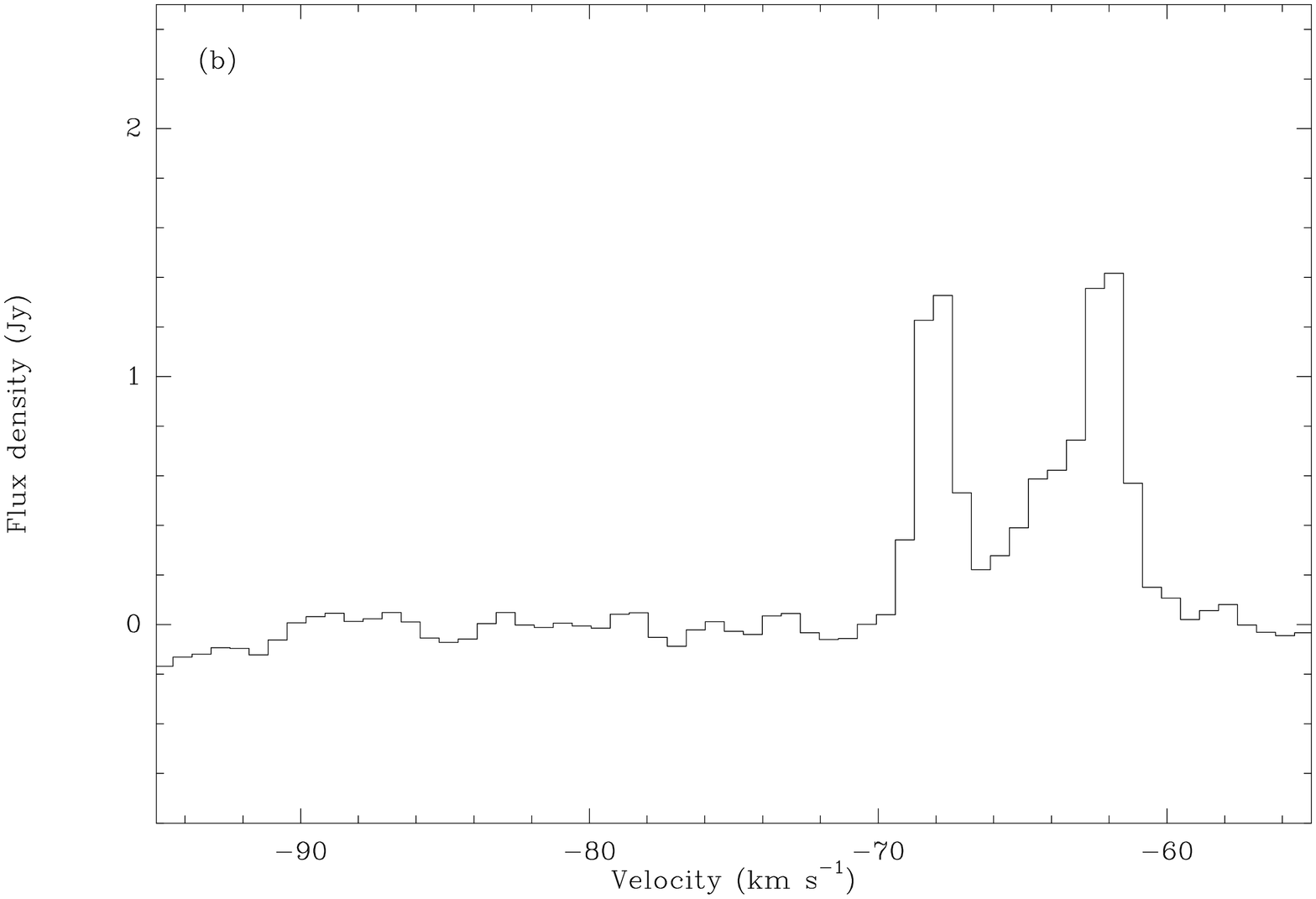}}
\caption{Water maser spectra of IRAS 17347-3139 (a) obtained with the
  Robledo antenna on day 131 of 2002, and (b) with the VLA (integrated
  over a region of $1\farcs 3\times 2\farcs 3$, which includes all the
  detected maser components) on day 189 of 2002.} \label{ispec}
\end{figure}

Table \ref{tbsurvey} shows the result of our survey of water maser emission towards
a sample of 27 planetary nebulae. Emission was detected in two sources
of the sample. One of them, K 3-35, 
was previously known to harbor a water maser, being
the first confirmed case of a bona-fide planetary nebula showing this
kind of
emission (MGAT01). We have also obtained a new detection
towards IRAS 17347-3139 with the Robledo antenna, on day 131 of 2002
(and confirmed with
the VLA, on day 189 of 2002, see Fig. \ref{ispec}), which
could represent the second case of a planetary nebula associated 
with water maser
emission.

\subsection{IRAS 17347-3139}

\subsubsection{Radio continuum emission}

\begin{figure}
\rotatebox{-90}{
\epsscale{0.6}
\plotone{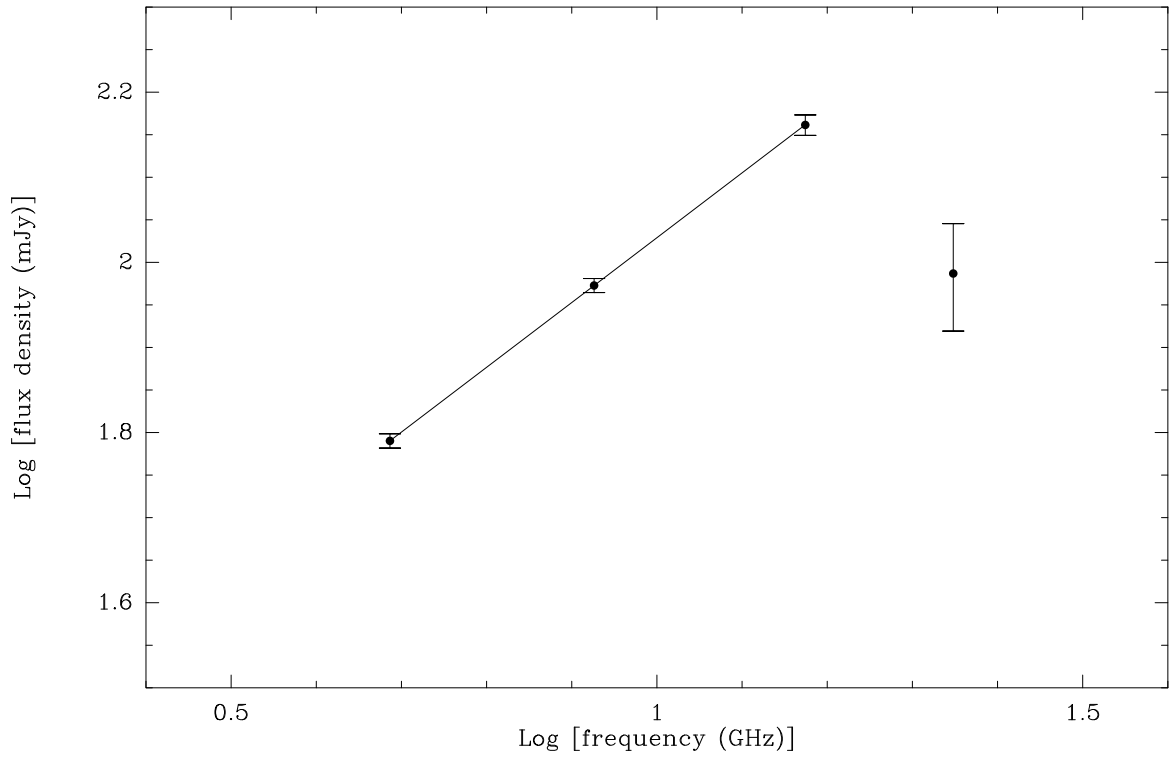}}
\caption{Log-log plot of flux density of IRAS 17347-3139 as a function of frequency,
  including $2\sigma$ error bars. The solid line is a linear fit
  between 4.86 and 14.94 GHz,
  which gives the spectral index.} \label{spindx}
\end{figure}

Radio continuum emission associated with IRAS 17347-3139 
is detected at all four observed
frequencies, unresolved at 4.9, 8.4, and 14.9 GHz, and marginally resolved
at 22.3 GHz (deconvolved size $\sim 0\farcs 3$).
This continuum emission is located at 
$\alpha({\rm J}2000) = 17^h 38^m 00\fs 586$,
$\delta({\rm J}2000) = -31\degr 40\arcmin 55\farcs 67$ 
(estimated from the 22.3 GHz image, 
absolute position error $\sim 0\farcs 25$).
 From the measured flux densities (Table \ref{tbl-4}), we derive a
spectral index ($S_\nu \propto 
\nu^\alpha$) of $\alpha = 0.76 \pm 0.03$ between 4.9 and 14.9 GHz  
(see Fig. \ref{spindx}),
indicating a thermal nature for this radio source. The
measured 22.3 GHz flux density is a factor of two lower than its
extrapolation using this spectral index ($97\pm 14$ versus $\sim 200$
mJy). Part of this loss of flux density could be due to source
variability, since  the 22.3 GHz data were observed at a
different epoch. Another possibility is that the 22.3 GHz
observations, being obtained with a more extended VLA configuration
than at lower frequencies, resolve out weak, extended emission due to
the lack of short spacings in the interferometer, thus resulting in a
net decrease of flux density. We also note that these data were 
observed at low elevation, which makes the 
absolute flux calibration at this wavelength more uncertain. However,
it is also likely that a significant 
fraction of the difference can be explained
by  the
spectral index becoming flat between 14.9  and 22.3 GHz. This
could indicate that the continuum emission becomes optically thin
between these frequencies. Quasi-simultaneous, matching-beam
observations around this 
frequency range would be needed to test this possibility, ruling out
variability, resolution, and calibration effects.

Extended and weaker radio continuum emission, located $\sim 45''$
southeast from
the unresolved source, is also detected in the 4.9 and 8.4 GHz maps. With
such a
large separation, this
extended emission is not likely to be related to IRAS 17347-3139.

\subsubsection{Water maser emission}

\begin{figure}
\rotatebox{-90}{
\epsscale{0.9}
\plotone{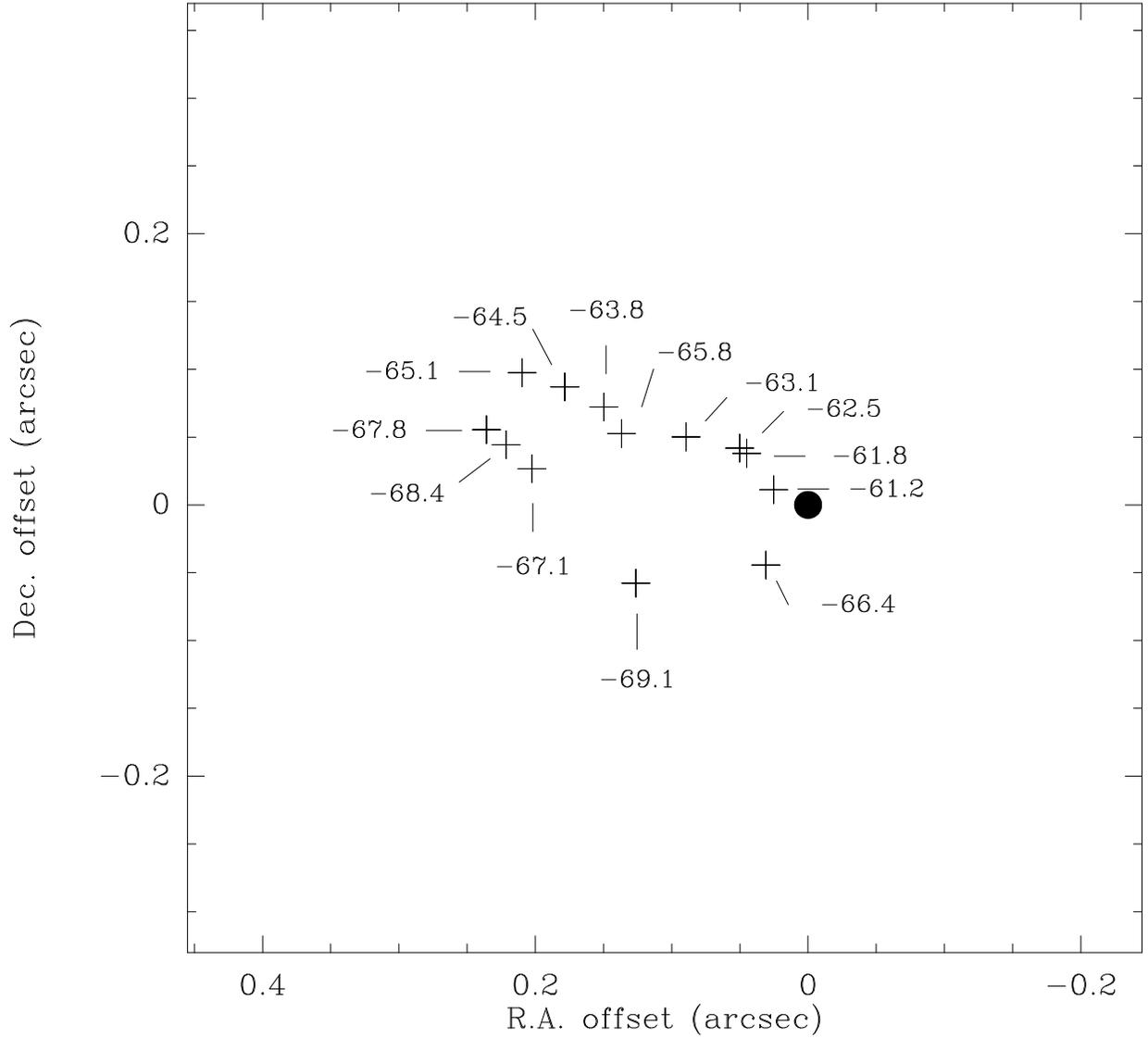}}
\caption{IRAS 17347-3139 water maser positions. Crosses mark the positions of
  the VLA maser components. The filled circle marks the position of
  the 1.3-cm continuum emission peak $[ \alpha({\rm J}2000) = 17^h 38^m 00\fs 586,
  \delta({\rm J}2000) = -31^\circ 40' 55\farcs67 ]$, which is the reference
  position in 
  this map (relative position error between masers and continuum $\sim
  0\farcs 03$). The beam size is $0\farcs 81 \times 0\farcs 26$ (P.A = 26$\degr$)} 
\label{ivla} 
\end{figure}

Thirteen water maser spots were detected with the VLA on day 189 of
2002 (see Table \ref{tbcomp} and
Fig. \ref{ivla}).
The spectrum of the whole
maser emission, integrated over an area of $1\farcs 3\times 2\farcs
3$,
 can be seen
in Fig. \ref{ispec}b.
The maser spots 
trace an elliptical structure, with the radio continuum peak near one of the
tips of its major axis. A least-square fit of the maser positions 
to an ellipse gives axes 
of $\simeq 0\farcs 25\times0\farcs 12$, 
with its major axis oriented at PA $\simeq 62^{\circ}$. 
 From the axis ratio of the ellipse, if we assume that this ellipse 
is tracing a circular ring, the axis of the ring would be oriented at $\sim
60^{\circ}$ with respect to the line of sight.

In addition, 
an analysis of the variation of the radial velocities in the ellipse 
reveals some trends. If we assume that the mean value of the observed 
velocities ($- 65.1$ km s$^{-1}$) represents the ``systemic velocity'' of 
the maser spots, 
the maximum (redshifted) radial velocities of 
$\simeq$ 3-4 km s$^{-1}$ tend to cluster towards the west of the maser
distribution, 
while the maximum (blueshifted) velocities of $\simeq$ 3 km s$^{-1}$
are at the east. 
This would suggest the presence of
rotational motions, although the kinematic trend is not compatible
with pure rotation, since the maximum velocities are not exactly at
the tips of the major axis of the ellipse. A combination of both
rotation and expansion in a
toroidal structure is necessary to explain this kinematic trend (see
\S\ref{ring}).

\subsection{K 3-35}
\begin{figure}
\rotatebox{-90}{
\plotone{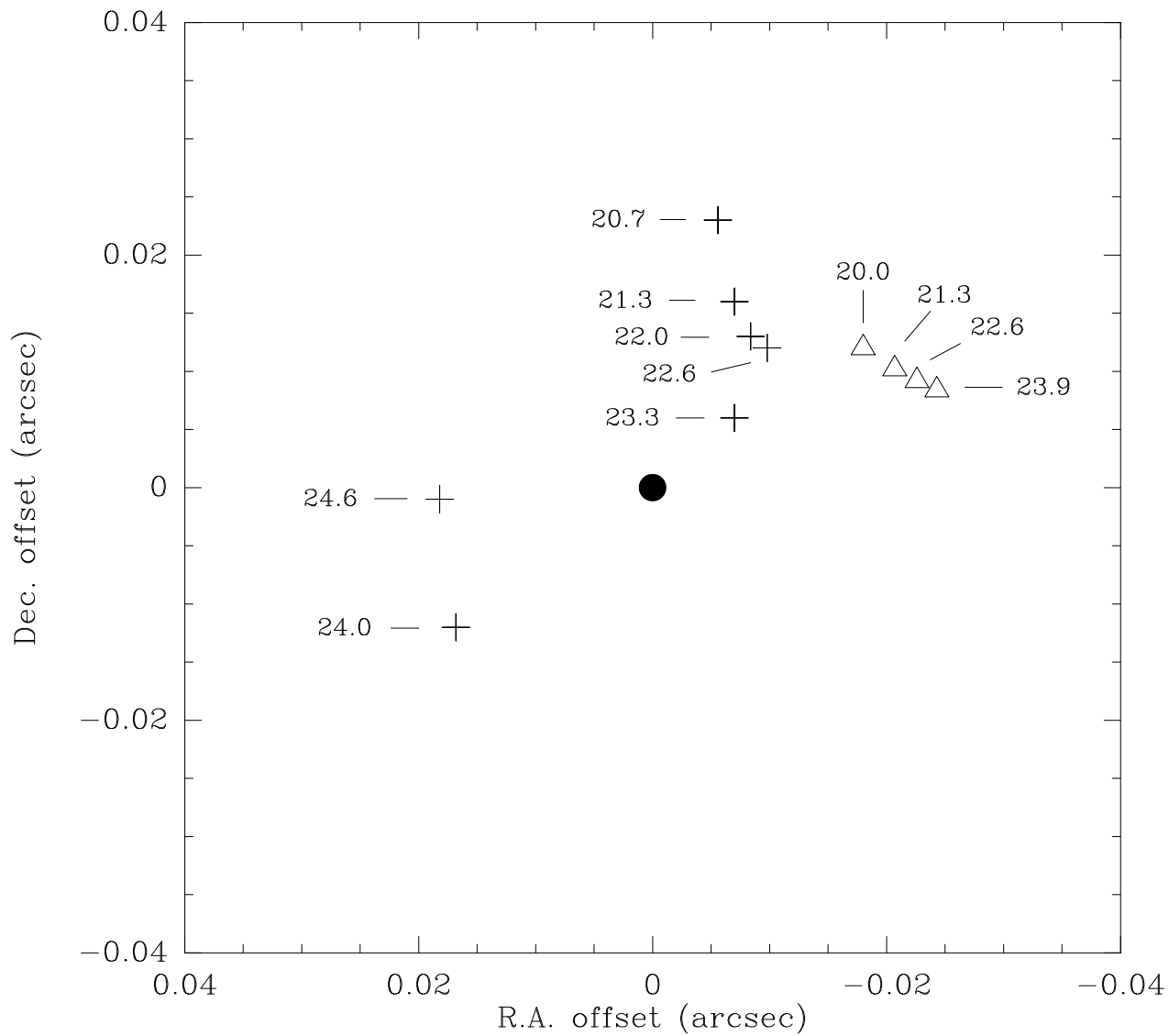}}
\epsscale{1.0}
\caption{K 3-35 water maser distribution. Triangles mark the positions of
  the maser components obtained in our new VLA observations, crosses
  mark the positions obtained in MGAT01 (see G\'omez et al 2003). The LSR 
velocity (in km s$^{-1}$) of each component is shown. The filled circle marks the position of
  the 1.3-cm continuum emission peak, assumed equal for both sets of
  observations, and it is the reference position for this map.
Relative position errors between
maser components and continuum emission are $\sim 0\farcs 004$ for our new
observations, and $\sim 0\farcs 001$ for those of MGAT01. The beam size
is $0\farcs 16 \times 0\farcs 09$ (P.A = 52$\degr$).} \label{kvla}
\end{figure}

With the VLA, 
we have detected four maser spots at the central region of this
source (see Table \ref{tbcomp} and Fig \ref{kvla}). 
In order to compare the positions of the two sets of water maser 
observations
(MGAT01 and this paper), we have obtained a continuum map at 1.3 cm
by averaging 51 channels free of water maser emission.
The 
peak of the continuum emission is located at 
$\alpha({\rm J}2000) = 19^h 27^m 44\fs 025$,
$\delta({\rm J}2000) = 21\degr 30\arcmin 03\farcs 43$
(absolute position error $\simeq 0\farcs 05$), 
which is consistent, within the errors, with the position  reported by
MGAT01. 
The two sets of water masers where then aligned with respect
to the 1.3 cm continuum peak position, assuming it coincident for the
two epochs, since the relative positional accuracy between continuum and
masers within each map ($\leq 0\farcs 004$) 
is better than their absolute position
accuracy ($\sim 0\farcs 05$). However, we cannot discard the presence
of small variations in the ionized structure of the source, that can
produce slight changes in the position of the continuum peak. 
The  overlay (Fig. \ref{kvla}) shows that 
the position and velocity distributions 
of the maser spots in 2002 are different from those observed in 1999 by 
MGAT01, which is not surprising giving the typical variability of this kind 
of emission. In particular, no maser emission was detected at the tips
of the bipolar lobes of the PN at a 3$\sigma$ level of 7 mJy, which
was instead 
detected in MGAT01. The maser spots observed in 2002 seem to correspond 
to the northwestern masers of the central 
region C detected in 1999, but shifted by $\sim 0\farcs 015$ toward the
west. 
However, the time difference between 
the two epochs ($\simeq$ 2.5 yr) is too long to make an unambiguous 
identification of the same water maser spots in the two epochs. 
Therefore a derivation of proper motions is probably meaningless. 
The 
observed positional variations would also be compatible with destruction 
of the water maser shell zone observed in 1999 by an ionization/shock front 
and the creation of a new maser zone ahead of the shock. New, more closely 
spaced multi-epoch observations, and with high S/N for the continuum
emission (to use it as an accurate position reference for the masers), 
are crucial to interpret the observed 
positional variations of the water maser spots in K 3-35.

\section{Discussion}

\subsection{The lifetime of water molecules in PN}

Two out of 27 PNe observed present water maser emission. This low rate
of detection seems to indicate that water molecules are destroyed in a 
very short time after the star enters the PN phase, or that
the excitation conditions for maser emission cease rapidly. 
The destruction
time of water molecules in PNe 
has been estimated to be $\sim 100$ yr \citep{Gom90}. 
It is difficult to make a quantitative statistical study since we do not
know the exact age range of our sample (the selection criteria were of
qualitative nature and the sample is probably biased towards young PNe), 
and the sensitivities of the 
telescopes used are significantly different. Nevertheless, if we
assume that our sample 
of PNe is homogeneously distributed in
age, along the 
$\sim 2\times 10^4$ yr that the PN phase lasts, our observed detection
rate of 1 out of 26 would be compatible with the estimated $\sim 100$
yr survival time of water molecules. 

In the following 
we will concentrate the discussion on IRAS 17347-3139, whose maser
emission is reported for the first time in this paper.

\subsection{IRAS 17347-3139}

\subsubsection{The distance to IRAS 17347-3139}

As far as we know, there is no published information about the
distance to this source. A crude estimate for this distance can be
obtained taking into account that water maser shells in red giants are located at a radius of $\simeq$ 10--100 AU (e.g.,
Spencer et al. 1979; Bowers et al. 1993). In addition, there
is evidence that the water maser shell moves outwards as
the star evolves and the mass loss rate increases (Yates \&
Cohen 1994). Therefore, if the angular radius of $0\farcs 12$ of
the water maser ring in IRAS 17347-3139 corresponds to a linear radius
of $\geq 100$ AU, a distance of $\geq 0.8$ kpc is obtained for
the nebula. 

As an independent estimate, 
from the observed flux density at 4.8 GHz and assuming a size of 
$0\farcs 3$ (the deconvolved size of the 22.3 GHz image) for the
ionized region, we can use  
the statistical distance scale for galactic PN 
proposed by \citet{Zha95}. We obtain
with this method a
value of $\sim 0.8$ kpc, which is consistent with the lower limit
mentioned above.

\subsubsection{The nature of IRAS 17347-3139 as a planetary nebula}

\label{pnsection}

IRAS 17347-3139 is a source with a large infrared excess
\citep{Zij89}.  
It shows OH maser emission \citep{Zij89} but no SiO maser emission
\citep{Nym98}. 
Its IRAS two-color diagram indicates that it is an evolved object 
\citep{Gar97}. However, this diagram cannot discriminate between proto-PNe 
and PNe. The classification of IRAS 17347-3139 as a PN was based on
the presence of  
radio continuum emission attributed to an ionized nebula \citep{Zij89}. Before 
this classification is conclusively established, other possibilities to 
explain the radio continuum emission should be ruled out.  

Strong radio continuum emission at centimeter wavelengths is observed in the 
\objectname{Egg Nebula}, a well known proto-PN, although its  spectral
index (2.6), too large to be due to free-free emission, suggests dust  
as the origin of the radio continuum emission \citep{Jur00}. However, in IRAS
17347-3139,  
the spectral index of 0.8, being lower than 2, rules out dust as
the origin of the radio continuum 
emission. 

A second possibility to explain the radio continuum emission
would be 
an ionized wind from the central star. In order to test this possibility, we
estimated the mass-loss rate required to explain the observed radio
continuum emission as arising from an ionized wind, which can be
compared with typical values of central stars of proto-PNe and PNe. 
Assuming a spherical wind with a constant velocity of 1000
km s$^{-1}$ and an electronic temperature of 10000 K, from the observed
flux density at 3.6 cm (Table 4) we obtain a value for the ionized
mass-loss rate of $\dot M_i\simeq 1\times10^{-4}~(D/\rm
kpc)^{3/2}~M_\odot$~yr$^{-1}$ (Wright \& Barlow 1975, Panagia \& Felli
1975). For the same set of parameters, collimated winds would require
smaller mass-loss rates.  For example, a biconical wind (corresponding to
a spectral index of 0.6) would require a mass-loss rate ten times
smaller, for an angle of aperture of $\sim 30^\circ$, and a collimated
wind with a variable angle of aperture (corresponding to the observed
value of the spectral index of $\sim 0.8$ between 4.9 and 14.9 GHz) a
mass-loss rate three times smaller than a spherical wind (see Reynolds
1986). Therefore, in all cases, the derived mass-loss rates are much larger than 
those observed in central 
stars of PNe and proto-PN ($\leq 10^{-7}$ M$_{\odot}$\,yr$^{-1}$, Patriarchi \&
Perinotto 1993; Vassiliadis \& Wood 1994), which
seems to exclude that the radio continuum emission arises in an ionized 
stellar wind.

All this suggests that the radio continuum emission in IRAS
17347-3139 arises in 
a partially optically thick ionized nebula and that it
is probably a very young PN. If we assume that the radio continuum
emission is optically thin for frequencies above $\sim 20$ GHz, we
derive an electronic density of $\sim 1\times 10^6$ cm$^{-3}$ for 
an homogeneous HII region at a distance of 0.8 kpc. The ionizing photon rate 
required to mantain such an HII region is $\dot N \simeq 10^{46}$ 
s$^{-1}$, that would correspond to a star with  T$_{\rm eff}\ga$
26\,000 K,  
according to \citet{pan73}. This temperature is higher than the T$_{\rm eff} \simeq$ 15\,000\,K deduced by
\citet{Zij89}, but it is more typical of young PNe. 

Finally, this source presents also infrared lines of [\ion{Ne}{2}] and
Br$\alpha$ (P. Garc\'{\i}a-Lario 2003, private communication), which confirms its
identification as a PN.

\subsubsection{Relationship between masers and infrared emission}

\begin{figure}
\rotatebox{-90}{
\epsscale{0.6}
\plotone{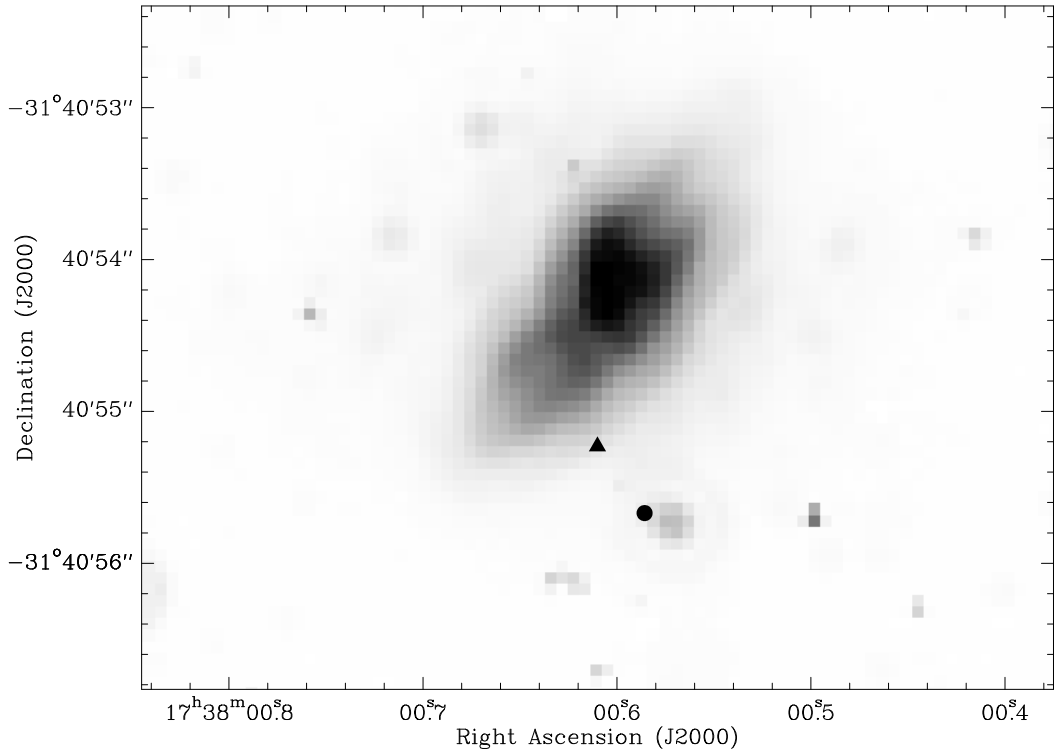}}
\caption{Infrared image (K band) of IRAS 17347-3139 taken with the Hubble
  Space Telescope. Grey levels are arbitrary, and have been chosen to
  show the main structure of the object. Absolute coordinates of the
  HST image are probably in error by $\sim 1''$. The filled circle shows the
  nominal position of the 22.3 GHz radio continuum (positional
  uncertainty $0\farcs 25$. The triangle marks
  the position of the 2MASS infrared point source (positional
  uncertainty $0\farcs 06$)}  \label{ihst}
\end{figure}

It is interesting to compare the water maser data with the K-band 
image of the object, taken with the NICMOS instrument of the Hubble
Space Telescope. We have retrieved this image from the 
HST archive and it is shown in Fig. \ref{ihst}. 
A preliminary description can be found 
in \citet{Bob99}. 
In addition to a faint envelope of $\simeq 3\farcs 5$ in size, the 
brightest regions of this source trace a clear bipolar structure 
with a size of $\simeq 2\farcs 0\times 0\farcs 7$ and the main nebular axis 
oriented at PA $\simeq -40^{\circ}$. The NW lobe is brighter than the SE 
one and they are separated by a dark lane that suggests the presence of an 
equatorial torus-like structure. 

We also note that there is a positional shift between the
nominal infrared position of the nebula given by the HST and the peak 
position
of the radiocontinuum emission, the latter being displaced 
$\sim$ 1$''$ south from the
center of the nebula. This angular separation might cast some doubts
on the relationship of the radiocontinuum and the 
water masers with the infrared
nebula. However, the 2MASS catalog lists an infrared point source at
$\alpha({\rm J}2000) = 17^h 38^m 00\fs 610$,
$\delta({\rm J}2000) = -31\degr 40\arcmin 55\farcs 23$ 
(absolute position error $\sim 0\farcs 06$), 
which is  $\sim 0\farcs 5$ northeast of the radio continuum
position, i.e., closer to the latter than the HST image center. An
offset of $\sim 0\farcs 5$ is reasonable even if it is larger than the
formal radio continuum position error ($0\farcs 25$), since we do not expect
the maximum of the unresolved infrared 2MASS image to exactly
coincide with the radio emission, taking into account that the
resolved HST image has its maximum in the northern part of the nebula.
Therefore, the discrepancy with the HST position is probably due to
errors in the  HST Guide Star Catalog, which can be of the order of $1''$.
Moreover, the association between the radio and infrared emissions is
reinforced by two further arguments: 
(1) The thermal nature of the radio source (see \S 4),
which seems to exclude the presence of a background extragalactic source,
and (2) the distribution of the water maser spots, with
the major axis of the ellipse oriented almost perpendicular to the
main nebular axis. This indicates a close relationship
and that the water maser emission could arise
in the dense equatorial ring related to the dark lane observed in the 
K-band
infrared image.

\subsubsection{The ring of water masers}
\label{ring}

As we mentioned above, to explain the kinematics of the water masers,
we need to assume the presence of both rotation and expansion in a
ring. 
The
uncertainty in the
distance to the source makes it difficult to estimate the velocity
component due to Keplerian rotation. If we assume that the expansion
and rotation velocity are of the same order, we obtain a reasonable
fit for a
torus radius of $0\farcs 12$, with a velocity of 2.4 km s$^{-1}$. For
such a rotational velocity, the central mass should be $M_*= 0.78
D$ M$_\odot$, where $D$ is the distance from the Sun to the source, in
kpc. This implies a central mass of $\sim 0.6$ M$_\odot$, assuming a distance
of 0.8 kpc. The water maser ring could trace the innermost regions of
the torus-like density enhancement formed at the very end of the AGB
evolution of the precursor star of IRAS\,17347-3139. 

As already mentioned, the radio continuum emission peak is located out 
of the ellipse traced by the water maser spots. 
An explanation for this could be the presence of 
inhomogeneities 
in the ionized region. Nevertheless, we also speculate on
the possible nature of IRAS 17347-3139 as a binary system, so
that the 
position of the continuum peak represents the {\it true}
central star, while
the water 
maser ellipse is associated with a companion. Our observations have not
enough resolution to confirm the binary nature of this source, since
the center of the water maser ellipse and the continuum peak are
separated a distance of $\sim 0\farcs15$, smaller than the beam size (see Fig. \ref{ivla}). This interpretation might
be confirmed by resolving the ionized region, which 
would allow us to determine its small-scale geometry, and whether a
double source is indeed present.

The water maser emission in both K 3-35 (see MGAT01) and IRAS
17347-3139 (this paper) are suggestive of
equatorial ring-like  
structures which have been formed at the end of the AGB phase. 
In the case of IRAS 17347-3139, the kinematics of the ring suggests
the presence of both rotating and expanding  
motions, while the observations of K 3-35 are compatible with only expansion 
in the equatorial plane. In addition, the effective temperature 
 for the central star of 
K 3-35 should be $\geq$ 60\,000\,K, as indicated by the He {\sc ii} emission 
\citep{Mir00}, larger than that obtained for the central star of 
IRAS 17347-3139 ($\ga$ 26\,000 K). Therefore, it is possible that IRAS
17347-3139 
represents an earlier stage in PN 
formation than K 3-35. The fast wind in IRAS 17347-3139 
has not yet been able to 
sweep up enough material in the (dense) equatorial plane and the water maser 
zone still preserves in part its original kinematics.

\section{Conclusions}

In this paper, we present a survey for water maser emission towards a
sample of 27 planetary nebulae, using the antennas at
Robledo de Chavela, and Medicina, as well as the VLA. Our main
conclusions are as follow:

\begin{enumerate}
\item Besides the known water maser in K 3-35 (MGAT01), we have detected
  a new source of maser emission 
towards IRAS 17347-3139 using the Robledo antenna (and
  later confirmed it with the VLA).
\item The low rate of new detections (1 out of 26) is compatible with the
  assumption that water molecules are rapidly destroyed ($\sim 100$
  yr) after the star
  enters the PN phase. 
\item Water masers in IRAS 17347-3139, observed with the VLA, show an
  elliptical distribution.
  This spatial distribution, and their radial velocities are suggestive
  of a rotating and expanding maser ring, 
tracing the innermost regions of a torus
  formed at the end of the AGB phase.
\item The continuum emission in IRAS 17347-3139 seems to arise from a 
  partially optically thick ionized region, since the mass
  loss rate obtained under the assumption of a wind 
is much larger
  than the values observed in PNe and proto-PNe.
\item We speculate on the possible nature of IRAS 17347-3139
  as a binary star, where the peak of the radio 
  continuum emission represents one of the
  components, and the maser ellipse is associated with a companion.
\end{enumerate}

\acknowledgments

GA, IdG, JFG, LFM, and JMT acknowledge support from MCYT grant (FEDER funds) 
AYA2002-00376 (Spain). JFG is also
supported by MCYT grant AYA 2000-0912. 
IdG acknowledges the support of a Calvo Rod\'es
Fellowship from the Instituto Nacional de T\'ecnica Aeroespacial. YG
acknowledges support from DGAPA-UNAM and CONACyT, Mexico. We
are thankful to Jes\'us Calvo, Cristina Garc\'{\i}a, Tom Kuiper,
Esther Moll, 
Pablo P\'{e}rez, and 
the operators at MDSCC for their
help before and during the observations at Robledo, and to Olga
Su\'arez and Pedro Garc\'{\i}a-Lario for their useful comments on the
manuscript. 
This paper is partly based on observations taken during
``host-country'' allocated time at Robledo de Chavela; this time is managed
by the LAEFF of INTA, under agreement with NASA/INSA.
It also makes use of data products from the Two Micron All 
Sky Survey, which is a joint project of the University of
Massachusetts and the Infrared Processing and Analysis 
Center/California Institute of Technology, funded by the 
National Aeronautics and Space Administration and the National Science
Foundation. We have also used 
observations made with the NASA/ESA Hubble Space Telescope,
obtained from the data archive at the Space Telescope Science Institute;
STScI is operated by the Association of Universities for Research in
Astronomy, Inc. under NASA contract NAS 5-26555.

\clearpage

\clearpage

\clearpage

\clearpage

\clearpage

\clearpage

\begin{deluxetable}{lrrrrrrrl}
\tabletypesize{\scriptsize}
\tablecaption{Water maser survey. \label{tbsurvey}}
\tablewidth{0pt}
\tablehead{
\colhead{Source} & \colhead{Right Ascension\tablenotemark{a}}   &
\colhead{Declination\tablenotemark{a}}   &
\colhead{$V_{\rm LSR}$\tablenotemark{b}} &
\colhead{$V_{\rm min}$\tablenotemark{c}} & 
\colhead{$V_{\rm max}$\tablenotemark{c}} &
\colhead{Flux density\tablenotemark{d}}   &
\colhead{DoY\tablenotemark{e}}     & 
\colhead{Telescope}\\
\colhead{} & \colhead{(J2000)} & \colhead{(J2000)} & \colhead{(km
  s$^{-1}$)} & 
\colhead{(km s$^{-1}$)} & \colhead{(km s$^{-1}$)} & 
\colhead{(mJy)} & \colhead{(of 2002)} & \colhead{}
}

\startdata

\objectname{PK 130-10.1} & 01 42  19.90 & 51 34 37.0 &   (-21.5)&
-122.7 & +79.7 & $<$400 &  225& Robledo\\              
\objectname{K 3-94} &           03 36 08.10&   60 03 46.6&  (-71.0)&
-111.5 & -30.5 & $<$11&   125&  VLA\\
\objectname{IC 2149} &          05 56 23.90&   46 06 17.0&  (-23.6)&
-76.2 & +28.9 & $<$900&   114& Medicina\\
   &                  &               &   (-38.0) & -78.5 & +2.5 &  
    $<$11&   125&  VLA\\
\objectname{NGC 2440} &         07 41 55.40&   -18 12 31.0&   (+52.2)&
0.2  & +104.2 &    $<$2200&    275& Medicina\\
   &                  &               &  (+39.6) &  -13.3
   &  +92.6 &    $<$1400&  300& Medicina\\
\objectname{NGC 4361} &         12 24 30.90&   -18 47 05.0&   (+20.0)&
-81.2 & +121.2 &   $<$300&       107& Robledo\\
   &                  &               &  (+20.0)&   -81.2 & +121.2 &
   $<$230&      108& Robledo\\ 
   &                  &               &  (+33.6)  &  -19.8 & +86.9 &
   $<$5000&    178& Medicina\\ 
   &                  &               &  (+27.0)  &  -20.1 &  +74.2 &
   $<$1200&    276& Medicina\\ 
   &                  &               &  (+19.6)  &  -34.0 &  +73.1 &
   $<$1400&    300& Medicina\\ 
\objectname{M 2-9}&            17 05 37.80&    -10 08 32.0&
(+101.4)&  +47.8 &  +154.9 &  $<$1400&     81& Medicina\\
   &                  &               &  (+87.9)  &   +34.9 & +140.9 &
   $<$1300&    301& Medicina\\ 
\objectname{IRAS 17106-3046} & 17 13 51.70&    -30 49 40.0&
(0.0)&  -101.2 & +101.2 &   $<$1000&      218& Robledo\\
\objectname{NGC 6309} &         17 14 03.60&    -12 54 37.0&
(-33.8)&  -87.6 &  +20.0 & $<$1700&   109& Medicina\\
&                     &              &   (-34.0)  &   -87.4 &  +19.4 &
$<$11000&   179& Medicina\\ 
&                     &              &   (-40.0)  &   -86.9  & +7.0 &
$<$1100&    276& Medicina\\ 
&                     &              &   (-47.1)  &   -100.9 &  + 6.7
&  $<$1400&    301& Medicina\\ 
\objectname{IRAS 17207-2856} & 17 23 55.80&    -28 59 32.0&
(-18.0)& -119.2 & +83.2 &   $<$1000&     218&  Robledo\\
\objectname{IRAS 17347-3139} & 17 38 00.70&  -31 40 54.0&
-62.1$\pm$0.5 &  -226.2 & +81.2 & 2100$\pm$300&    131&  Robledo\\ 
        &            &               &          -68.0$\pm$0.5& && 2100$\pm$300&    131&  Robledo\\ 
        &            &               &          -68.1$\pm$0.5& -168.2
        & +34.2 &  700$\pm$400&    140&  Robledo\\ 
        &            &               &          -61.8$\pm$1.3&  -95.7
        & -54.3 & 1360$\pm$60\phn&    189&  VLA\\
        &            &               &          -68.2$\pm$0.5&  -166.2
        & +36.2 & 1700$\pm$600&    218&  Robledo\\ 
\objectname{IRAS 17375-3000} & 17 40 44.80&    -30 02 00.0&
(-25.0)&  -126.2 & +76.2 & $<$1200&       218& Robledo\\ 
\objectname{IRAS 17423-1755} & 17 45 14.20&    -17 56 47.0&
(+35.0)& -66.2 & +136.2 &  $<$190&       108& Robledo\\ 
        &            &               &   (+35.0)   &   -66.2 & +136.2
        &    $<$800&       218& Robledo\\  
\objectname{NGC 6572} &         18 12 06.28&   06 51 12.6&  (-10.0)&
-111.2 & +91.2 & $<$110&        110& Robledo\\
        &            &               &   (-10.0)   &   -50.5 & +30.5 &
        $<$12&      125& VLA\\ 
\objectname{Cn 3-1} &           18 17 34.09&   10 09 03.4&  (+38.0)&
-63.2 & 139.2 & $<$140&       110& Robledo\\ 
        &            &               &   (+38.0)   &   -2.5 & +78.5 &
        $<$13&      125& VLA\\ 
\objectname{M 3-27} &           18 27 48.27&   14 29 06.2&  (-24.0)&
-64.5 & +16.5 & $<$13&      125& VLA\\
\objectname{M 2-43} &           18 26 40.10&    -02 42 57.0&
(0.0)&  -101.2 & +101.2 &     $<$140&        108&  Robledo\\
        &            &               &       (0.0)&  -101.2 & +101.2
        &      $<$300&        167&  Robledo\\  
        &            &               &      (+11.8) &  -40.9 &  +64.4 &
        $<$1200&      274& Medicina\\ 
\objectname{M 1-57} &           18 40 20.30&    -10 39 47.0&
(+92.0)& -9.2 & 193.2 &  $<$110&       108&  Robledo\\
        &            &               &       (+92.0)& -9.2 & 193.2
        &      $<$130&        131&  Robledo\\  
\objectname{Hu 2-1}&           18 49 47.56&   20 50 39.4&  (+46.8)&
-6.5  & +100.0 &  $<$900&        81& Medicina\\
        &            &               &     (+33.0) &  -7.5 & +73.5 &    $<$13&     125& VLA\\
\objectname{IC 4846}&          19 16 28.23&   -09 02 36.6&   (-168.0)&
-269.2 & -66.8 & $<$100&        108& Robledo\\ 
        &            &               &       (-168.0)&
-269.2 & -66.8       &      $<$130&        131&  Robledo\\
        &            &               &    (-156.2) &   -208.8 & -103.5
        &   $<$1700&      274& Medicina\\ 
\objectname{Vy 2-2} &           19 24 22.20&   09 53 56.0&  (-64.0)&
-104.5 & -23.5 & $<$14&      125& VLA\\
\objectname{K 3-35} &           19 27 44.02&   21 30 03.4&
22.4$\pm$0.5 &  -76.2 & +126.2 & 270$\pm$50\phn&        108& Robledo\\  
        &            &               &           22.6$\pm$1.3& -20.5 &
        +60.5 & 933$\pm$5\phn\phn & 125& VLA\\
        &            &               &           22.4$\pm$0.5&   -76.2
        & +126.2 &  1330$\pm$90\phn&        131& Robledo\\
        &            &               &           22.4$\pm$0.5&   -76.2
        & +126.2 & 1410$\pm$130&        140& Robledo\\ 
        &            &               &           22.4$\pm$0.5&   -76.2
        & +126.2 & 1330$\pm$120&        246& Robledo\\ 
        &            &               &           22.4$\pm$0.5&   -76.2
        & +126.2 & 2100$\pm$400&       284&  Robledo\\

\objectname{M 2-48} &           19 50 28.00&   25 54 28.0&   (0.0)&
-40.5 & +40.5 &    $<$13&       125&  VLA\\
\objectname{NGC 6884} &         20 10 23.59&   46 27 39.4&  (-21.0)&
-61.5 & +19.5 & $<$12&    125&  VLA\\
\objectname{NGC 6881} &         20 10 52.40&   37 24 42.0&  (-14.0)&
-115.2 & +87.2 &  $<$150&       110&  Robledo\\
        &            &               &    (­14.0)   &   -54.5 & +26.5
        &   $<$13&  125&  VLA\\ 
\objectname{IC 4997} &          20 20 08.80&   16 43 53.0&  (-66.0)&
-106.5 & -25.5 & $<$15&      125&  VLA\\
\objectname{IC 5117} &          21 32 30.90&   44 35 47.0&  (0.0)&
-101.2 & +101.2 &     $<$200&        131&  Robledo\\
        &            &               &    (0.0)    &  -40.5 & +40.5  &  $<$12&       125&  VLA\\
\objectname{KjPn 8} &           23 24 10.45&   60 57 30.5&  (-21.8) &
-75.4 &   +31.8&   $<$1000&    109&  Medicina\\
        &            &               &      (-35.0)       &  -75.5 &
        +5.5 &     $<$12&    125&  VLA\\

\enddata
\tablenotetext{a}{Coordinates of phase center in VLA observations, or
  pointing position of single-dish data. Units of right ascension are
  hours, minutes, and seconds. Units of declination are degrees,
  arcminutes, and arcseconds}
\tablenotetext{b}{Velocity of the strongest detected maser
  components. For non-detections,  the central velocity of the observed bandwidth is given
  between parentheses}
\tablenotetext{c}{Velocity range covered by the observational bandwidth}
\tablenotetext{d}{For non-detections, upper limits are $3\sigma$. For
  detections, uncertainties are $2\sigma$}
\tablenotetext{e}{Observation date, day of year in 2002}

\end{deluxetable}

\clearpage

\begin{deluxetable}{lrrrr}
\tablecaption{Phase calibrators and synthesized beams of VLA water
  maser observations\tablenotemark{a}\label{tbcals}}
\tablewidth{0pt}
\tablehead{
\colhead{Source} & \colhead{Phase calibrator}   & \colhead{Flux
  density\tablenotemark{b}}   &
\colhead{Beam size\tablenotemark{c}} &
\colhead{Beam P.A.\tablenotemark{c}} \\
\colhead{} & \colhead{} & \colhead{(Jy)}&  \colhead{} & \colhead{}
 
}

\startdata
K 3-94&      J0244+624&   1.23$\pm$0.05\phn&     $0\farcs 10 \times 0\farcs 07$&  36$\degr$\\
IC 2149&     J0555+398&   2.62$\pm$0.12\phn&     $0\farcs 11 \times 0\farcs 07$&  57$\degr$\\
IRAS 17347-3139& J1745$-$290& 0.986$\pm$0.008&    $0\farcs 81 \times 0\farcs 26$&  26$\degr$\\
NGC 6572&    J1849+005&   0.723$\pm$0.003&   $0\farcs 16 \times 0\farcs 09$ &  45$\degr$\\
Cn 3-1&      J1849+005&   0.723$\pm$0.003&   $0\farcs 16 \times 0\farcs 09$&  48$\degr$\\ 
M 3-27&      J1849+005&   0.723$\pm$0.003&   $0\farcs 16 \times 0\farcs 09$&  53$\degr$\\
Hu 2-1&      J1925+211&   2.979$\pm$0.130&   $0\farcs 15 \times 0\farcs 09$&  56$\degr$\\
Vy 2-2&      J1950+081&   0.334$\pm$0.014&   $0\farcs 16 \times 0\farcs 09$&  48$\degr$\\
K 3-35&      J1925+211&   2.979$\pm$0.130&   $0\farcs 16 \times 0\farcs 09$&  52$\degr$\\
M 2-48&      J2023+318&   0.634$\pm$0.025&   $0\farcs 13 \times 0\farcs 09$&  59$\degr$\\
NGC 6884&    J2012+464&   0.485$\pm$0.018&   $0\farcs 11 \times 0\farcs 09$& -85$\degr$\\
NGC 6881&    J2015+371&   3.56$\pm$0.11\phn&     $0\farcs 13 \times 0\farcs 09$&  73$\degr$\\
IC 4997&     J2031+123&   0.529$\pm$0.017&   $0\farcs 18 \times 0\farcs 10$&  56$\degr$\\
IC 5117&     J2202+422&   2.33$\pm$0.08\phn&     $0\farcs 11 \times 0\farcs 09$&  90$\degr$\\
KjPn 8&      J2322+509&   0.679$\pm$0.020&   $0\farcs 11 \times 0\farcs 08$& -45$\degr$\\

\enddata
\tablenotetext{a}{All sources were observed with A configuration,
  except IRAS 17347-3139 that was observed with B configuration}
\tablenotetext{b}{Bootstrapped flux density of the phase calibrator}
\tablenotetext{c}{Size and position angle of the synthesized beams of
the images}
\end{deluxetable}

\clearpage

\begin{deluxetable}{rr}
\tablecaption{IRAS 17347-3139 continuum emission \label{tbl-4}}
\tablewidth{0pt}
\tablehead{
\colhead{Frequency} &\colhead{Flux density\tablenotemark{a}} \\
\colhead{(GHz)} &
\colhead{(mJy)} \\

}
\startdata
4.86& 61.67$\pm$1.2\\ 
8.44& 93.9$\pm$1.8\phn\\
14.94&  145$\pm$4\phd\phn\phn\\ 
22.29&  97$\pm$14\phd\phn\\
\enddata
\tablenotetext{a}{Uncertainties are  $2\sigma$}
\end{deluxetable}

\clearpage
\begin{deluxetable}{lrrrrl}
\tabletypesize{\scriptsize}
\tablecaption{Water maser components detected with VLA \label{tbcomp}}
\tablewidth{0pt}
\tablehead{
\colhead{Source} & 
\colhead{       $V_{\rm LSR}$\tablenotemark{a}} &
\colhead{Flux density\tablenotemark{b}} & \colhead{Right Ascension}   &
\colhead{Declination}  & \colhead{Position
  uncertainty\tablenotemark{b,c}}  \\
\colhead{} & \colhead{(km s$^{-1}$)} & \colhead{(mJy)} &
\colhead{(J2000)} & \colhead{(J2000)} & \colhead{($\arcsec$)}  

}

\startdata
IRAS 17347-3139&    -61.2&   440$\pm$50&  17 38 00.589\phn\phn& 
-31 40 55.66\phn\phn &   0.04  \\
        &           -61.8&  1310$\pm$60&  17 38 00.5903\phn&  -31 40
55.632\phn&   0.017  \\
        &           -62.5&  1360$\pm$60&  17 38 00.5907\phn&  -31 40
55.628\phn&   \nodata\tablenotemark{d}  \\
        &           -63.1&   720$\pm$50&  17 38 00.594\phn\phn&   -31 40
55.62\phn\phn&   0.03  \\
        &           -63.8&   540$\pm$50&  17 38 00.598\phn\phn&  -31 40
55.60\phn\phn&   0.04  \\
        &           -64.5&   410$\pm$50&  17 38 00.601\phn\phn&  -31 40
55.58\phn\phn&   0.05  \\
        &           -65.1&   260$\pm$50&  17 38 00.603\phn\phn&  -31 40
55.57\phn\phn&   0.07  \\
        &           -65.8&   180$\pm$50&  17 38 00.597\phn\phn&  -31 40
55.62\phn\phn&   0.10  \\
        &           -66.4&   150$\pm$50&  17 38 00.589\phn\phn&  -31 40
55.71\phn\phn&   0.13  \\
        &           -67.1&   330$\pm$50&  17 38 00.603\phn\phn&  -31 40
55.64\phn\phn&   0.06  \\
        &           -67.8&  1020$\pm$50&  17 38 00.6052\phn&  -31 40
55.615\phn&   0.021  \\     
        &           -68.4&  940$\pm$50&  17 38 00.6041\phn&  -31 40
55.626\phn&   0.022  \\
        &           -69.1&   220$\pm$50&  17 38 00.597\phn\phn&  -31 40
55.73\phn\phn&   0.09  \\
K 3-35   &            23.9&   351$\pm$6\phn&   19 27 44.02296&  21 30
03.4404&  0.0012  \\             
        &            22.6&   933$\pm$5\phn&   19 27 44.02308& 21 30
03.4412& \nodata\tablenotemark{e} \\
        &            21.3&   714$\pm$6\phn&   19 27 44.02322& 21 30
03.4422& 0.0006  \\
        &            20.0&   125$\pm$6\phn&   19 27 44.0234\phn&  21 30
03.444\phn&  0.004  \\
\enddata
\tablenotetext{a}{Velocity of water maser emission}
\tablenotetext{b}{Uncertainties are $2\sigma$}
\tablenotetext{c}{Relative position uncertainties with respect to the
  reference feature used for self-calibration}
\tablenotetext{d}{Reference feature. Absolute position error is $\sim
  0\farcs 25$}
\tablenotetext{e}{Reference feature. Absolute position error is $\sim
  0\farcs 05$}
\end{deluxetable}


\begin{thebibliography}{}
\bibitem[Aaquist (1993)] {Aaq93} Aaquist, O.~B. 1993, \aap, 267, 260. 
\bibitem[Aaquist \& Kwok (1989)] {AyK89} Aaquist, O.~B., \& Kwok, S. 1989, 
  \aap, 222, 227
\bibitem[Ali et al. (2001)]{Ali01} Ali, A., Shalabiea, O. M., El-Nawawy,
  M. S., \& Millar, T. J. 2001, MNRAS, 325, 881 
\bibitem[Balick (1987)] {Bal87} Balick, B. 1987, \aj, 94, 671. 
\bibitem[Bobrowsky, Greeley, \& Meixner (1999)] {Bob99} Bobrowsky, M., 
Greeley, B., \& Meixner, M. 1999, BAAS, 31, 1536
\bibitem[Bowers et al. (1993)]{Bow93} Bowers P.F., Claussen, M.J.,
  Johnston K.J. 1993, AJ, 105, 284
\bibitem[Elitzur (1992)] {Eli92} Elitzur, M. 1992, \araa, 30, 75. 
\bibitem[Engels (2002)] {Eng02} Engels, D. 2002, \aap, 388, 252
\bibitem[Garc\'{\i}a-Lario et al. (1997)] {Gar97} Garcia-Lario, P., 
  Manchado, A., Pych, W., \& Pottasch, S.~R. 1997, A\&AS, 126, 479
\bibitem[G\'omez et al. (2003)] {Gom03} G\'omez, Y., Miranda, L. F.,
  Anglada, G., \& Torrelles, J. M. 2003,  
in IAU Symp 209, Planetary
      Nebulae. Their Evolution and Role in the Universe,
      eds. S. Kwok, M. Dopita, \& R. Sutherland
      (San Francisco: ASP), in press
\bibitem[G\'omez et al. (1990)] {Gom90}  G\'omez, Y., Moran, J.~M., \& 
  Rodriguez, L.~F. 1990, RevMexAA, 20, 55.  
\bibitem[Habing (1996)] {Hab96} Habing, H.~J. 1996, \aapr, 7, 97.    
\bibitem[Hasegawa et al. (2000)]{Has00} Hasegawa, T., Volk, K., \&
  Kwok, S. 2000, ApJ, 532, 994
\bibitem[Jura et al. (2000)] {Jur00} Jura, M., Turner, J.~L., Van Dyk, S., \& 
  Knapp, G.~R. 2000, \apjl, 528, L105
\bibitem[Kwok (1993)] {Kwo93} Kwok, S. 1993, \araa, 31, 63
\bibitem[Lane et al. (1987)]{lane87} Lane, A. P., Johnston, K. J., Bowers,
  P. F., Spencer, J. H., \& Diamond, P. J. 1987, ApJ, 323, 756
\bibitem[Lewis (1989)] {Lew89} Lewis, B.~M. 1989, \apj, 338, 234. 
\bibitem[Likkel \& Morris (1988)] {Lik88} Likkel, L., \&  Morris,
  M. 1988, \apj, 329, 914. 
\bibitem[Marvel (1997)] {Mar97} Marvel, K.~B. 1997, \pasp, 109, 1286.   
\bibitem[Marvel \& Boboltz (1999)] {MyB99} Marvel, K.~B., \&  Boboltz,
  D.~A. 1999, \aj, 118, 1791.
\bibitem[Miranda et al. (2000)] {Mir00} Miranda, L.~F., Fern\'andez, M., 
  Alcal\'a, J.~M., Guerrero, M.~A., Anglada, G., G\'omez, Y., Torrelles, 
  J.~M., \& Aaquist, O.~B. 2000, \mnras, 311, 748
\bibitem[Miranda et al. (2001)] {Mir01} Miranda, L.~F., G\' omez, Y.,
  Anglada, G., \& Torrelles, J.~M. 2001, \nat, 414, 284 (MGAT01)
\bibitem[Miranda et al. (1998)] {Mir98} Miranda, L.~F., Torrelles, J.~M., 
  Guerrero, M.~A., Aaquist, O.~B., \& Eiroa, C. 1998, \mnras, 298, 243 
\bibitem[Nyman, Hall, \& Olofsson (1998)] {Nym98} Nyman, L.-A., Hall, P.~J., \&
  Olofsson, H. 1998, A\&ASS, 127, 185
\bibitem[Panagia (1973)]{pan73} Panagia, N. 1973, \aj, 78, 929
\bibitem[Panagia \& Felli (1975)] {PyF75} Panagia, N., \& Felli, M. 1975, 
  \aap, 39,1
\bibitem[Patriarchi \& Perinotto (1993)] {PyP93} Patriarchi, P., \&
  Perinotto, M. 1993, A\&ASS, 325, 335
\bibitem[Reid \& Moran (1981)]{RyM81} Reid, M.~J. \& Moran,
  J.~M. 1981, \araa, 19, 231
\bibitem[Reynolds (1986)]{rey86} Reynolds, S. P. 1986, ApJ, 304, 713
\bibitem[Sahai \& Trauger (1998)] {SyT98} Sahai, R., \& Trauger, J.~T.
  1998, \aj, 116, 1357
\bibitem[Sahai et al. (1999a)] {Sah99a} Sahai, R., Te Lintel Hekkert,
  P., Morris, M., Zijlstra, A., Likkel, L. 1999a, \apj, 514, L115
\bibitem[Sahai et al. (1999b)] {Sah99b} Sahai, R., Zijlstra, A.,
  Bujarrabal, V., Te Lintel 
  Hekkert, P. 1999b, \aj, 117, 1408
\bibitem[Spencer et al. (1979)]{Spe79}
Spencer J.H., Johnston K.J., Moran J.M., Reid M.J.,
Walker R.C. 1979, ApJ, 230, 449
\bibitem[Vassiliadis \& Wood (1994)]{Vas94} Vassiliadis, E., \& Wood,
  P. R. 1994, ApJS, 92, 125
\bibitem[Wright \& Barlow (1975)] {WyB75} Wright, A.~E., \& Barlow, M.~J.
  1975, \mnras, 170, 41 
\bibitem[Yates \& Cohen (1994)]{Yat94}
Yates J.A., Cohen R.J., 1994, MNRAS, 270, 958
\bibitem[Zhang (1995)] {Zha95} Zhang, C.~Y., 1995, \apjs, 98, 659 
\bibitem[Zijlstra et al. (1989)] {Zij89} Zijlstra, A.~A. te Lintel Hekkert, P.,
  Pottasch, S.~R., Caswell, J.~L., Ratag, M., \& Habing, H.~J. 1989, \aap, 217, 
  157


\end{thebibliography}
\end{document}